\documentstyle[12pt]{article}
\title{A Possible Explanation of the CP Puzzle}

\author{ Georgij Takhtamyshev}
\begin{document}
\maketitle
 
\begin{center}
{\it Centro Brasileiro de Pesquisas F\'\i sicas, Rio de Janeiro
 22290-180,  Brasil}\\
{\it and}\\
{\it Joint Institute for Nuclear Research, Dubna 141980, Russia}\\

\end{center}

\begin{abstract}
  The problem of mirror-reflection symmetry (MRS) and time-reversal
symmetry (TRS) in our world is discussed.

 The opinion is expressed, that well-known  experiments
on parity violation and CP-violation can be treated as signals of
some new, yet unknown, level of matter.

  An hypothesis, which can be used as a base for some future model
or theory is formulated.
  In the framework of this hypothesis,  experiments which
demonstrate parity violation or CP-violation do not contradict
MRS or TRS conservation.

\end{abstract}


  The hypothesis of possible parity violation was suggested in
1956 by Lee and Yang \cite{LY} and then experimentally confirmed
by  Wu with collaborators in 1957 \cite{Wu}.
 CP-violation was experimentally discovered in 1964 by Christenson,
Cronin, Fitch and Turlay \cite{CP}.
 Since then, these phenomena remain unsolved puzzles in
modern physics.
  Very probably at present, especially for the younger generation
of physicists, some witnesses are needed to confirm that violation
of P and CP are puzzling phenomena.
The very direct and open opinion of Feynman about parity violation
can be found in his book \cite{Fn}.
  And very nice sketch of events following discovery of CP-violation
is presented by Cronin in his Nobel lecture \cite{Nob}.

 Speaking here about MRS I  will use the definition given in
Ref. \cite{VW}. Namely I assume that  the  symmetry
exists if the following rule is valid:

{\em  The probability for any process equals the probability for
the mirror image of that process.}

  Similarly, speaking about TRS, I assume it exists if:

{\em The probability for any process
equals the probability for the time-reversed image of that process.}

 The general opinion is that present experimental and theoretical 
results contradict MRS conservation.
   A typical statement, reflecting this general opinion
 can be found, for example, in the book of R.Sachs \cite{Sachs}.

 But there is another view 
in which parity still can be considered an exact symmetry.

 We start with a short history of  exactly conserved parity.
 The first general statement about conserving parity
was expressed by Lee and Yang
in Ref. \cite{LY}. They said that parity may still be conserved if some
partners of the usual particles exist which exhibit opposite asymmetry.
 Wigner \cite{Wig1} suggested an hypothesis, according to which
each antiparticle is an exact mirror copy of particle. In this case the
experiment of Wu does not contradict MRS and parity is conserved.
But the main psychological barrier to accepting this idea is that 
electric charge in this case is pseudoscalar, not scalar \cite{Wig2}.
 In spite of lack of experimental data about the internal structure of 
electric charge, the general opinion is that it is a scalar \cite{WP}.

Some possibilities for explanation of the experimental data without
violation of left-right symmetry were considered by Yu.Shirokov
\cite{Shi}. He claimed that a particle under spatial inversion
and temporal reversal is transformed into some other state (e.g.
into an antiparticle).

  But it became clear after 1964 that antiparticles are not mirror
copies of particles, hence, some other candidates for mirror
counterparts of usual particles are needed for parity
conservation theory.
  Hypothetical mirror particles were described at first in a paper of
Kobzarev, Okun and Pomeranchuk in 1966 \cite{MP}.
According to their hypothesis each usual  particle  has  a  mirror
twin with the same mass. Mirror particles have their own  electric
charges and their own weak and strong interactions such that they
interact with particles of our world only gravitationally
and through  some new superweak field.
 But in  the  framework  of  their hypothesis, MRS
 still does not  exist.  A  weak  interaction  Lagrangian
remains both P- and CP-non-invariant,  but  now  it  is  invariant
relative to the CPA-transformation, where A is  a  new  operation,
transforming usual particles into mirror ones, and vice versa.
Respectively, all the charges are scalars.

 A very interesting example of parity
conservation theory was presented by M.Pavsic \cite{Pav}.
 His main idea is that all elementary particles have internal 
structure with internal spatial and temporal degrees of freedom. 
Hence to fulfil total space inversion ($P_{T}$), one has not only 
to make usual external inversion ($P_{E}$), but also internal 
inversion ($P_{I}$). In other words, for the total description of 
an elementary particle we have to introduce new variables 
corresponding to these internal degrees of freedom.
 Effectively, this is done by adding to the standard set of variables
(coordinate, momentum, spin) two new parameters $\alpha$ and $\tau$.
Parameter $\alpha$ has two discrete values: $+1$ and $-1$, and internal
spatial inversion changes the sign of this parameter.
 Parameter $\tau$ also has the same two discrete values $+1$ and $-1$,
and again internal time reversal changes the sign of this parameter.

 One more left-right symmetric theory was considered by Foot, Lew
and Volkas \cite{FLV}. The theory also includes mirror particles
with their own weak, strong, and electromagnetic interactions. 
The question about non-scalar charges is not discussed in the paper.

  It might be said that the difference between parity violating
(like, for example \cite{MP}), and parity conserving (\cite{FLV})
mirror particle theories is not very large. Namely, the difference 
is just in the name of the operator; what is named operator A in 
one paper is called operator P in another.

 But it should be emphasized that the difference is far from
being just formal; a parity conserving theory may lead us to
the unfolding of the internal structure of quarks and leptons.

Experiments demonstrating CP-violation conclude that the amplitude 
of the transition
${\it {\overline K^0} \rightarrow K^0}$
is greater than the same for the transition
${\it K^0 \rightarrow {\overline K^0}}$.
   This phenomenon does not contradict directly neither MRS nor
TRS validity. It only demonstrates some strange inequality between
matter and antimatter in our world.
  And, assuming validity of the CPT-symmetry, violation
of CP-symmetry means also violation T-symmetry. Then processes, 
demonstrating absence of TRS, like e.g. electric dipole momentum
can be observed. Taking into account that classic analog of spin
is rotation, one may say, that such thing as time-reversal image
of rotating neutron with electric dipole momentum never can be
observed.

  The point of interest is that P- and CP-violation phenomena 
can be considered as a manifestation of some new level of matter,
more primary than quarks and leptons.
Let's consider the mirror particle hypothesis, assuming that mirror
particles are true mirror images of usual particles.
 In this case the MRS exists in the 
exact sense of the word, and  electric and other charges can
not be scalars, since simple reflection of spatial axes transforms
each charge into another kind of charge.

 But, if charge changes its property at mirror reflection, it's
quite natural to assume that it has some space-time structure. 
Hence, some new, more fundamental level of matter
exists and our particles and antiparticles, together with their mirror
counterparts, are just made of the same pre-particles, or preons,
just as left and right molecules of sugar are made of 
the same atoms.
  And what we call electric and other charges is just the result 
of specific movement of these preons. The idea of electric charge
being the result of some movement is not new having appeared 
in early Kaluza-Klein models.

  At present nobody can suggest a quantitative
model of quarks and antiquarks made of these moving
preons, but one can point out some features of such a future  model.
There could be one or more types of preons, but none of them
possess any of the well-known charges (electric, weak, strong).
As a result of their interaction they make four different kinds of 
objects, namely, quarks,
antiquarks, mirror quarks, and mirror antiquarks. The system
representing a quark is the mirror copy of a system representing
a mirror quark and time reversal transforms both quark and mirror
quark into antiquark and mirror antiquark, respectively.

This last feature was already presented in the early
Kaluza-Klein model. Electric charge in the model arose as a result
of movement of the particle along the additional (compactified)
fourth space dimension. And the direction of this movement 
defined the charge of the particle. So the electron was exactly the 
same as a positron observed in the reversed time frame.

Usual and mirror particles and their
antiparticles can be described as states
with inner degrees of freedom (as it was done by Pavsic \cite{Pav}).
Let ${\it \Psi_1(x,\alpha,\tau)}$ and ${\it \Psi_2(x,\alpha,\tau)}$
be wave functions, describing particles and  antiparticles, whereas
wave functions ${\it \Psi_3(x,\alpha,\tau)}$ and 
${\it \Psi_4(x,\alpha,\tau)}$ 
describe mirror particles and antiparticles, respectively.
  All four of these functions are neither spatially, nor temporally
symmetrical. Instead of this, the following relations take place
\begin{equation}
   {\it \Psi_1(x,\alpha,\tau) = \Psi_3(x,-\alpha,\tau)}
\end{equation}
\begin{equation}
   {\it \Psi_2(x,\alpha,\tau) = \Psi_4(x,-\alpha,\tau)}
\end{equation}
\begin{equation}
   {\it \Psi_1(x,\alpha,\tau) = \Psi_2(x,\alpha,-\tau)}
\end{equation}
\begin{equation}
    {\it \Psi_3(x,\alpha,\tau) = \Psi_4(x,\alpha,-\tau)}
\end{equation}

 Evidently, if equations (1)--(4) are valid for all leptons and for
all quarks (both usual and mirror), then what we call P-violating
and T-violating phenomena look quite natural and understandable.
And inequality between matter and antimatter now is not strange,
because at the same time we observe symmetrical inequality
between mirror matter and mirror antimatter.

 Within this hypothesis such T-violating phenomenon as neutron 
(or any other particle) with electric dipole momentum does not
contradict TRS because observing it in the reversed time 
we see the same picture, as observing antineutron in the 
 direct time. 

 More generally, on the preon level all phenomena clearly
demonstrate the validity of both MRS and TRS. We conclude
that these symmetries seem to be violated in our experiments
only because so far
we have only observed two of the four kinds of composite systems. 
Physicists of the 19-th century could similarly claim a violation
of MRS when observing the rotation of the plane of linearly polarized
light passing through a sugar solution. As long as the second form
of the sugar molecule were not yet discovered, the quality of the proof 
of their statement would be precisely as good as the quality of proof 
of parity violation in the 20-th century.

  The main idea expressed in this paper can be stated in the 
following way:

{\em Experimental data demonstrating
CP-violation forces us to guess that there is something moving
inside quarks and leptons. And the only difference between a quark
and its respective antiquark is the direction of the movement of
this "something".}

  Any theory which incorporates this idea will be
free from the CP puzzle.

{\bf Acknowledgements.}
   The author is grateful to Dr.  V.  Nikitin (JINR) and  to  Dr.
D. Bardin (JINR) for very  helpful  discussions  and  support.
Discussions on the last stage of work on the paper with 
Dr. A. Demichev (Moscow State University)
were very fruitful and pleasant.


\begin{thebibliography}{99}

\bibitem{LY}  T. D. Lee, C. N. Yang,   Phys. Rev. {\bf 104}, 254 (1956)
\bibitem{Wu}  C. S. Wu {\it et al.,} Phys. Rev. {\bf 105}, 1413 (1957)
\bibitem{CP} 
 J. H. Christenson, J. W. Cronin, V. L. Fitch, R. Turlay,
 Phys. Rev. Lett.    {\bf 13}, 138 (1964)
\bibitem{Fn}
"This is a mystery that no one has the slightest idea about yet".

 R. P. Feynman, {\it The Character of Physical Law}
  (Penguin Books, 1992) p. 107.

\bibitem{Nob}
  "We must continue to seek the origin of the CP symmetry violation
by all means at our disposal. We know that improvements in detector
technology and quality of accelerators will permit even more sensitive
experiments in the coming decades.
 We are hopeful, then, that at some epoch, perhaps distant,
this cryptic message from nature will be deciphered."

 J. W. Cronin,  Rev. Mod. Phys. {\bf 53}  373 (1981).

\bibitem{VW}  K. Gottfried, V. Weisskopf, {\it Concepts of Particle 
Physics}  (Clarendon Press, Oxford, 1984) Vol. 1, p. 25.

\bibitem{Sachs}
  "Inversion enters into considerations of physics because of another
common perception - that right- and left-handedness are matters of
convention rather than of substance. Thus until Lee and Yang (1956)
proposed and Wu and others (1957) confirmed a contradiction to this
assumption, inversion was included among the transformations expressing
the isotropy of space, and right- and left-handed coordinate systems
were taken to be indistinguishable. Either implicitly or explicitly,
the dynamic laws were assumed to be invariant under inversion."

 R. G. Sachs, {\it The Physics of Time Reversal}
   (The University of Chicago Press, Chicago, 1989), p. 2.

\bibitem{Wig1}  E. Wigner, Rev. Mod. Phys. {\bf 29}, 255 (1957)

\bibitem{Wig2} The hypothesis of Wigner by no
  means coincides with the CP-conservation hypothesis. The
 hypotheses differ in the main point. 
 Wigner recognizes the antiparticle as the true mirror counterpart 
of the particle, while for the CP-conservation hypothesis
 the mirror counterpart of neutrino simply does not exist.
 In other words, Wigner is looking for an answer to the very 
relevant questions: "{\em Why} is CP conserved?
 {\em Why} does the operation C produce the same effect as
  the very different operation P ?".

\bibitem{WP}  It is
 just remarkable that Pauli has no doubt that 
 the origin of such a fundamental entity as electric charge can be
 clarified by some mysterious convention.

  "Nach der \"ublichen Auffassung (Konvention) \"andert bei der 
Operation P die elektrische Ladung ihr Vorzeichen nicht, so dass 
die elektrische Feldst\"arke ein axialer Vektor ist."

   W. Pauli, Experimentia vol. XIV, 1 (1958)

\bibitem{Shi} Yu. M. Shirokov, Zh. Eksp. Teor. Fiz. (Sov. Phys. JETP)
         {\bf 36}, 879 (1959) (in Russian)

\bibitem{MP}  I. Yu. Kobzarev, L. B. Okun, I. Ya. Pomeranchuk,
  Yad. Fiz. (Sov. J. Nucl. Phys.) {\bf 3}, 1154 (1966) (in Russian)

\bibitem{Pav} M. Pavsic, Int. J. Theor. Phys. {\bf 9}, 229 (1974)

\bibitem{FLV} R. Foot, H. Lew, R. R. Volkas, Phys. Lett.
    {\bf B272}, 67 (1991)

\end{thebibliography}
\end{document}